\documentclass{eas}
\usepackage{graphicx}
\newcommand{\LCDM}{$\Lambda$CDM}
\newcommand{\ML}{\ensuremath{\Upsilon_{\star}}}
\newcommand{\MLmax}{\ensuremath{\Upsilon_{max}}}
\newcommand{\MLpop}{\ensuremath{\Upsilon_{pop}}}
\newcommand{\MLopt}{\ensuremath{\Upsilon_{acc}}}

\newcommand{\Q}{\ensuremath{{\cal Q}}}
\newcommand{\Pop}{\ensuremath{{\cal P}}}
\newcommand{\G}{\ensuremath{{\Gamma}}}

\newcommand{\Md}{\ensuremath{{\cal M}_d}}

\newcommand{\Vb}{\ensuremath{V_{b}}}
\newcommand{\Vh}{\ensuremath{V_h}}
\newcommand{\Vf}{\ensuremath{V_{f}}}

\begin{document}

\TitreGlobal{Mass Profiles and Shapes of Cosmological Structures}

\title{Some Systematic Properties of Rotation Curves}
\author{McGaugh, S.}\address{Department of Astronomy,
	University of Maryland, College Park, MD, USA}
%
\runningtitle{Rotation Curves}
\setcounter{page}{23}
\index{McGaugh, S.}

\begin{abstract} 
The rotation curves of spiral galaxies obey strong scaling relations.
These include the Tully-Fisher and baryonic Tully-Fisher relations, and the
mass discrepancy---acceleration relation.
These relations can be used to place constraints on the mass-to-light
ratios of stars.  Once the stellar mass is constrained, the distribution of
dark matter follows.  The shape of the dark matter distribution is consistent
with the expectations of NFW halos exterior to 1 kpc, but the amplitude
is wrong.  This is presumably related to the long-standing problem
of the normalization of the Tully-Fisher relation and may imply a 
downturn in the amplitude of the power spectrum at small scales.  
More fundamentally, the persistent success of MOND remains a 
troubling fact.
\end{abstract}

\maketitle

\section{Introduction}

It is abundantly clear that there are mass discrepancies in the universe.  
Whether these imply the existence of invisible mass or a
modification of dynamical laws is less clear.
Either way, remarkable new physics must be involved.



The persistent flatness of rotation curves at radii well beyond the point
where the observed baryon distribution predicts a declining $V(R)$
was one of the key observations that convinced us of the need
for dark matter (e.g., Bosma 1981; Rubin \etal\ 1982).  While the
inference of dark matter is a perfectly natural one, strictly speaking
the data merely indicate the presence of mass discrepancies.
This can mean either dark matter or a breakdown in
the equations relating mass and dynamics.  Here I briefly review the
application of rotation curves as a test of each of these cases in turn.

\begin{figure}[h]
   \centering
   \includegraphics[width=9cm]{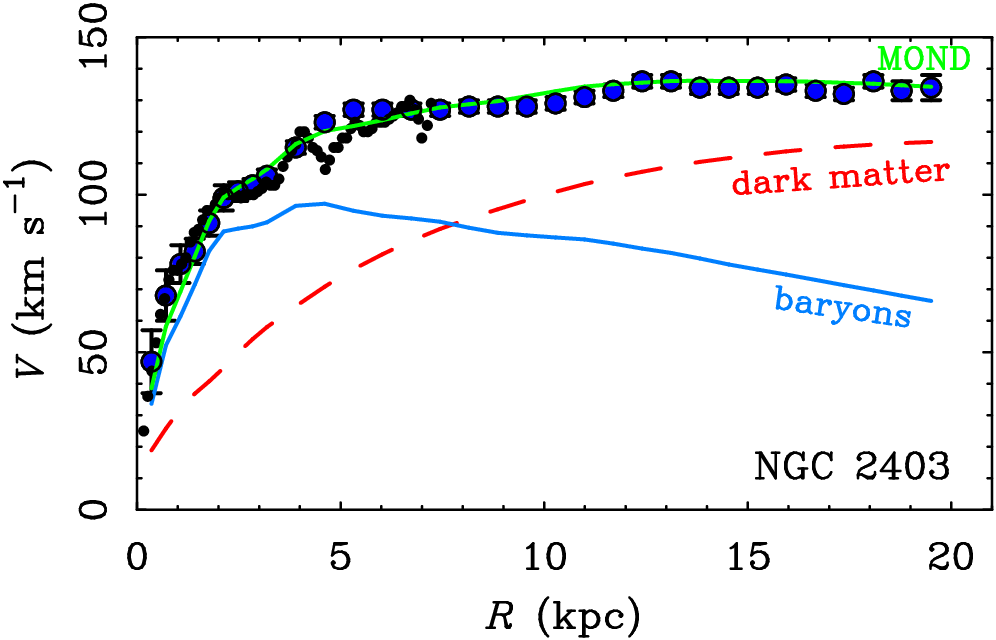}
      \caption{ Rotation curve and mass model of NGC 2403.  The large points
      are 21 cm data from Begeman \etal\ (1991); the small points
      are H$\alpha$ Fabry-Perot data from Blais-Ouellette \etal\ (2004).
      The declining solid line illustrates the contribution of the
      baryons (stars + gas); the dashed line that of the dark matter.
      These illustrate the combination of conventional components
      required to match the MOND fit (upper solid line).}
       \label{N2403}
   \end{figure}
  
In principle, rotation curves provide an excellent probe of the radial 
force law in galaxies.  
Fig.~\ref{N2403} shows an example of an extended rotation curve.
In practice, a number of issues complicate the
use of rotation curves to constrain the mass distribution.  These fall
into two broad categories:  issues concerning data quality (the accuracy
with which the potential is traced, e.g., Swaters \etal\ 2003; 
de Blok \etal\ 2003) and the degeneracy of modeling 
dark and baryonic components.  
While there are always cases for which the data can be improved,
the literature contains many excellent rotation curves that have been
consistently measured by independent observers using different techniques.
Yet even for perfect data, there is a terrible ambiguity between dark
and luminous mass.  The stellar mass is only known as well as the
mass-to-light ratio \ML, and the inferred dark matter distribution can
change dramatically for different \ML.  This effect is much stronger 
than the uncertainties in the rotation curve itself.

\section{The Tully-Fisher Relation and Stellar Mass-to-Light Ratios}

For any individual galaxy, the degeneracy between dark and luminous mass
is hard to lift.  However, galaxies as a population obey a number of scaling
relations (e.g., Persic \& Salucci 1988).  We can test a variety of prescriptions
for \ML; presumably the one that minimizes the scatter in relations
like the baryonic Tully-Fisher relation is the one that comes closest to the truth.

\begin{figure}[h]
   \centering
   \includegraphics[width=12cm]{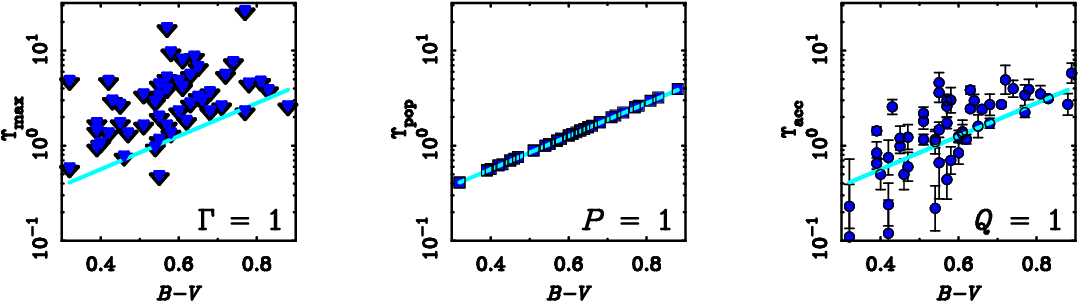}
      \caption{Stellar mass-to-light ratios as a function of $B-V$ color.
      \textbf{Left panel:} maximum disk. 
      \textbf{Middle panel:} stellar population synthesis (Bell \etal\ 2003).
      \textbf{Right panel:} \MLopt\ (MOND).  
      The population model of Bell \etal\ (2003) is shown as a line in all
      three panels.}
       \label{MLBV}
   \end{figure}

McGaugh (2005) considered a variety of prescriptions:
scaling \ML\ as a fraction of maximum disk
$\ML = \G \MLmax$; relative to population synthesis models
$\ML = \Pop \MLpop$; and relative to the mass discrepancy---acceleration
relation (the empirical version of MOND): $\ML = \Q \MLopt$.
These \ML\ are shown as a function of color in Fig.~\ref{MLBV}.

\begin{figure}[h]
   \centering
   \includegraphics[width=11cm]{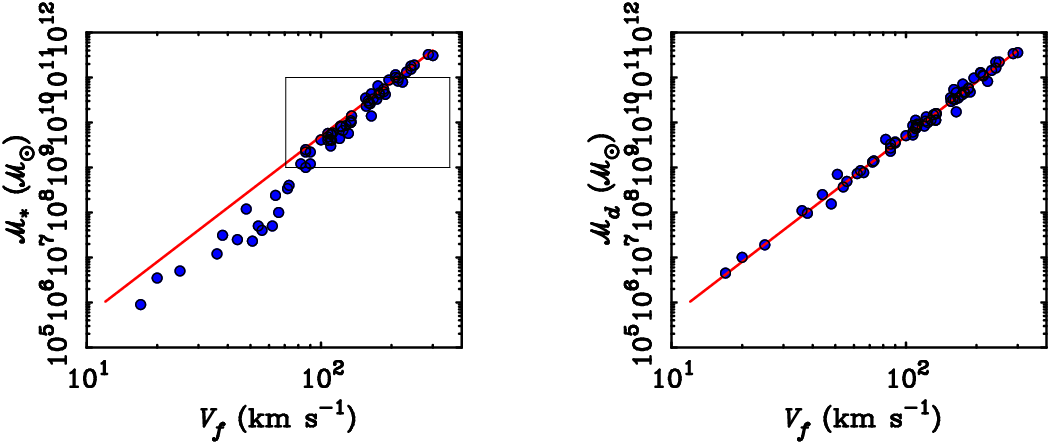}
      \caption{ \textbf{Left panel:} The stellar mass Tully-Fisher relation.
      \textbf{Right panel:} The baryonic Tully-Fisher relation (stars plus gas).
      The latter is well fit over five decades in mass by $\Md = 50 \Vf^4$ (line).
      This is a considerably larger dynamic range than has been available until
      recently.  The box in the left panel illustrates the range over which most
      Tully-Fisher work has been done.}
       \label{btf}
   \end{figure}

The prescription that is most consistent with stellar population synthesis
models is \MLopt.  Indeed, this is more consistent with such models
than the models are with themselves, in that a realistic amount of scatter 
is built in.  The case of $\Q =1$ also minimizes the scatter in the baryonic
Tully-Fisher relation (Fig.~\ref{btf}) and, by construction, in the 
mass discrepancy---acceleration relation.  The upshot of this is that 
\textbf{the stellar mass-to-light ratio is well determined}.

\section{The Dark Matter Distribution}

\begin{figure}[h]
   \centering
   \includegraphics[width=9cm]{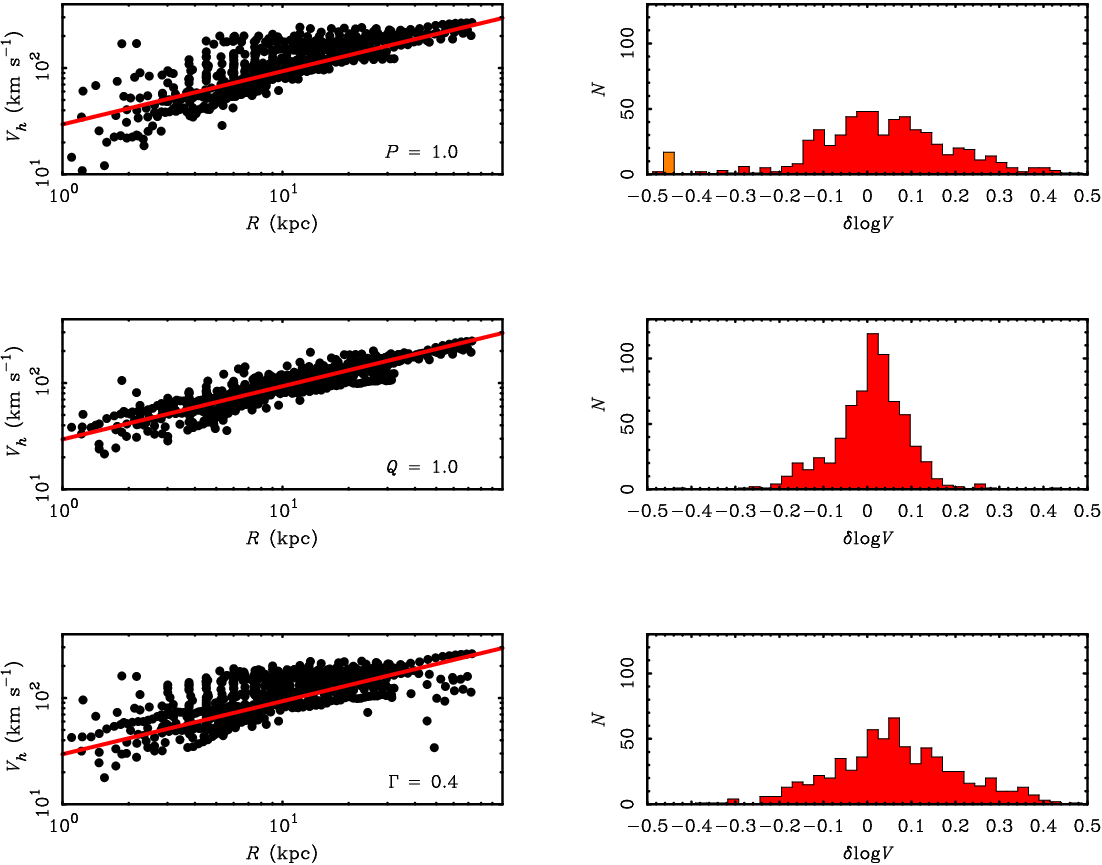}
      \caption{\textbf{Left panels:} The inferred rotation curves
      of the dark matter halos
      (equivalent to the dashed line in Fig.~\ref{N2403})
      of sixty galaxies (with over 600 individual points) 
      with high quality rotation curves (each point has a formal
      uncertainty of 5\% or less). 
      Each row illustrates a different prescription for the mass-to-light 
      ratios of the stars.  \textbf{Right panels:} the distribution of
      points around the solid line, which is the best fit to the case of
      minimum scatter (middle row).}
       \label{DMhisto}
   \end{figure}

Once \ML\ is determined, the dark matter distribution follows.
We need not limit consideration to the optimal case of $\Q = 1$.
For each prescription for \ML, we can ask what dark matter distribution
is inferred.  This is shown in Fig.~\ref{DMhisto}.

When examined in detail, there are clear differences between the
halo rotation curves of individual galaxies (Sancisi 2004; McGaugh 2004).
However, when plotted together as in Fig.~\ref{DMhisto}, one is struck
more by the over all similarity.  At a crude level, the data are consistent
with all galaxies living in identical dark matter halos.  Presumably, there
is some mass spectrum of halos, but that is not obvious from these data.

   \begin{figure}[h]
   \centering
   \includegraphics[width=9cm]{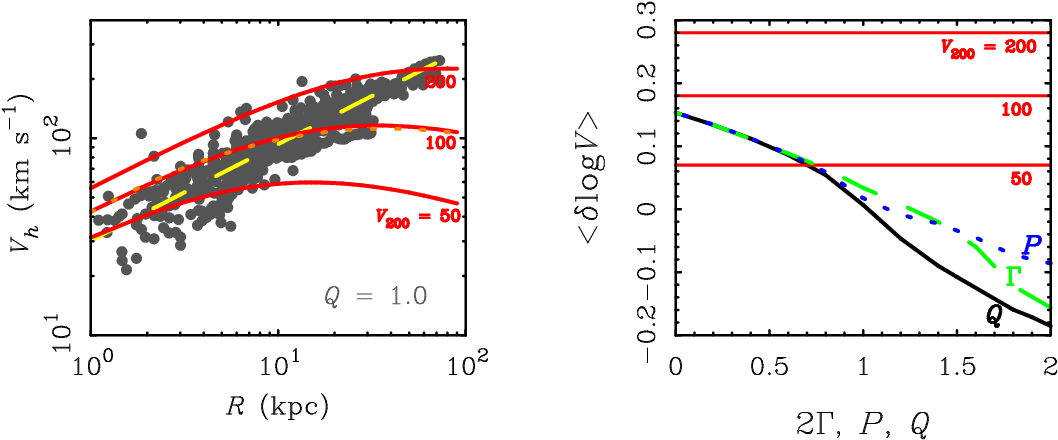}
      \caption{ \textbf{Left panel:}
      the inferred dark matter rotation curves 
      for sixty galaxies 
      tabulated in McGaugh (2005).
      The dashed line illustrates a slope of $1/2$, equivalent
      to a dark matter distribution whose density falls as $R^{-1}$.
      This is a good representation of the mean of the data for
      $R > 1$ kpc.  The solid lines show NFW halos for several
      $V_{200}$ with concentrations as predicted by the vanilla
      \LCDM\ parameters of Tegmark \etal\ (2004).
      Also shown (dotted line) is the profile suggested by
      Navarro \etal\ (2004), which is virtually indistinguishable
      from NFW.  \textbf{Right panel:}
      the variation of the mean velocity of the dark matter with
      different choices of mass estimator for the stars.
      The mean dark matter velocity increases as 
      stellar mass declines, but hardly intersects with the region predicted
      by \LCDM\ (horizontal lines).   }
       \label{VhQdV}
   \end{figure}

Remarkably, the scatter in the inferred rotation curves is minimized
for the case of $\Q = 1$.  The scatter in the dark matter component in this case
is comparable to that expected in CDM simulations (Bullock \etal\ 2001).  
A fit to these data yield a line
\begin{equation}
\log \Vh = A + B \log R
\end{equation}
with $A = 1.47$ and $B = 0.49 \pm 0.01$.
The value of the intercept $A$ depends on the choice of mass-to-light ratio,
as tracked by the centroid of the distribution in the right panels of 
Fig.~\ref{DMhisto}.  

It is striking that the slope $B \approx 1/2$.  This slope in velocity space
corresponds to a density distribution $\rho \propto R^{-1}$, which is the
inner slope of an NFW halo.  This sheds no light on the cusp/core controversy,
as the data here do not probe the inner kpc where that problem arises 
(de Blok 2004).  It does,
however, tell us about the density and structure of halos at intermediate radii.

What we observe is a continually rising halo rotation curve.
We do not see the flat portion of an isothermal halo:
the observed flat velocity is the combination of a declining
baryonic component and a rising dark halo component.
That \Vf\ remains roughly constant while its components
vary with radius [$\Vf^2 = \Vb^2(R) + \Vh^2(R)$] 
remains a curious puzzle.

The slope of the rising halo $\Vh(R)$ is consistent
with the expectation for an NFW halo.  However, the amplitude
is wrong, and the $B \approx 1/2$ slope persists rather too far.  
Over this range of radii, we expect to see significant curvature
in the halo rotation curves (Fig.~\ref{VhQdV}).
Presumably, galaxies reside in halos of a variety of different mass halos,
so perhaps we are merely seeing the locus of rising portions of the
halo rotation curves.  If this is the case, it is rather odd that we see
so much of the rise but so little of the turn-over.

The observed amplitude of $\Vh(R)$ poses a serious problem.
CDM predicts velocities that are too high.  This holds over the entire 
observed range of radii, irrespective of the detailed shape of the profile.

For vanilla \LCDM\ parameters (Tegmark \etal\ 2004), we expect
the dark matter velocity to be higher than observed by $\sim 0.25$ dex.
This is a large effect.  In terms of cosmological parameters,
the observations (for $\Q = 1$)
imply $\sigma_8 \Gamma_{0.6} \approx 0.05$,
where $\Gamma_{0.6} \propto \Omega_m^{0.6} h$ is a 
modified shape parameter (McGaugh, Barker, \& de Blok 2003).
In contrast, vanilla \LCDM\ has $\sigma_8 \Gamma_{0.6} = 0.24$.

Reconciling this difference would require 
large changes to basic cosmological parameters.  
This does not seem plausible.
It may be sufficient to suppress the power spectrum on small scales, 
but it is quite a stretch to maintain consistency with other constraints.  

The most obvious solution is to reduce the stellar mass ($\Q < 1$).
Fig.~\ref{VhQdV} shows how the centroid of the data varies with
different scalings of \ML.  The data just barely enter the region predicted
by \LCDM\ in the limit $\ML \rightarrow 0$, and then only
for rather lower mass halos than we typically associate with bright galaxies
(which are well represented in this sample).  We are not at liberty to
decrease \ML\ this much, as constraints on the IMF no longer
permit arbitrarily small \ML.  Moreover, the scatter in the baryonic
Tully-Fisher and mass discrepancy---acceleration relations becomes
unacceptably large for $\Pop < 1/2$ (McGaugh 2004, 2005).
Consequently, no matter how one chooses to look at the problem,
and irrespective of the details of the slope of the halo (cusp or core),
\textbf{the velocity attributable to dark matter
is significantly smaller than predicted by \LCDM.}

\section{MOND and the Mass Discrepancy---Acceleration Relation}

\begin{figure}[h]
   \centering
   \includegraphics[width=9cm]{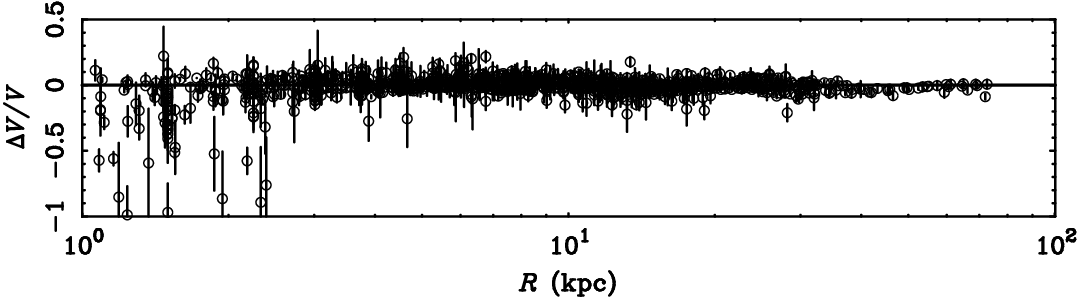}
      \caption{ Residuals of the MOND fits for 74 galaxies reviewed by
      Sanders \& McGaugh (2002).  Data for all galaxies are plotted together.
      MOND fits to rotation curves are generally good.  There is some tendency 
      for MOND to over-fit the velocity at small radii, which can plausibly be 
      attributed to non-circular motions and measurement difficulties 
      (e.g., Swaters \etal\ 2003; de Blok \etal\ 2003).
      These are far smaller than what would be required to reconcile the
      data with the dense NFW halos predicted by \LCDM.}
       \label{resid}
   \end{figure}

In stark contrast to the situation for CDM, MOND fits rotation curves well
(Fig.~\ref{resid}).  That is does so is well established 
(Sanders \& McGaugh 2002).  What this means is open to debate.

Like most astronomers, I ignored MOND for a long time, thinking it so
unlikely as to not warrant consideration.  I was obliged to reconsider when
the fine-tuning problems with dark matter became severe (McGaugh \&
de Blok 1998a), and the predictions of MOND
came true in my data.  Milgrom (1983) made a series of
specific predictions for low surface brightness galaxies, all of which
were subsequently realized (McGaugh \& de Blok 1998b).

\begin{figure}[h]
   \centering
   \includegraphics[width=9cm]{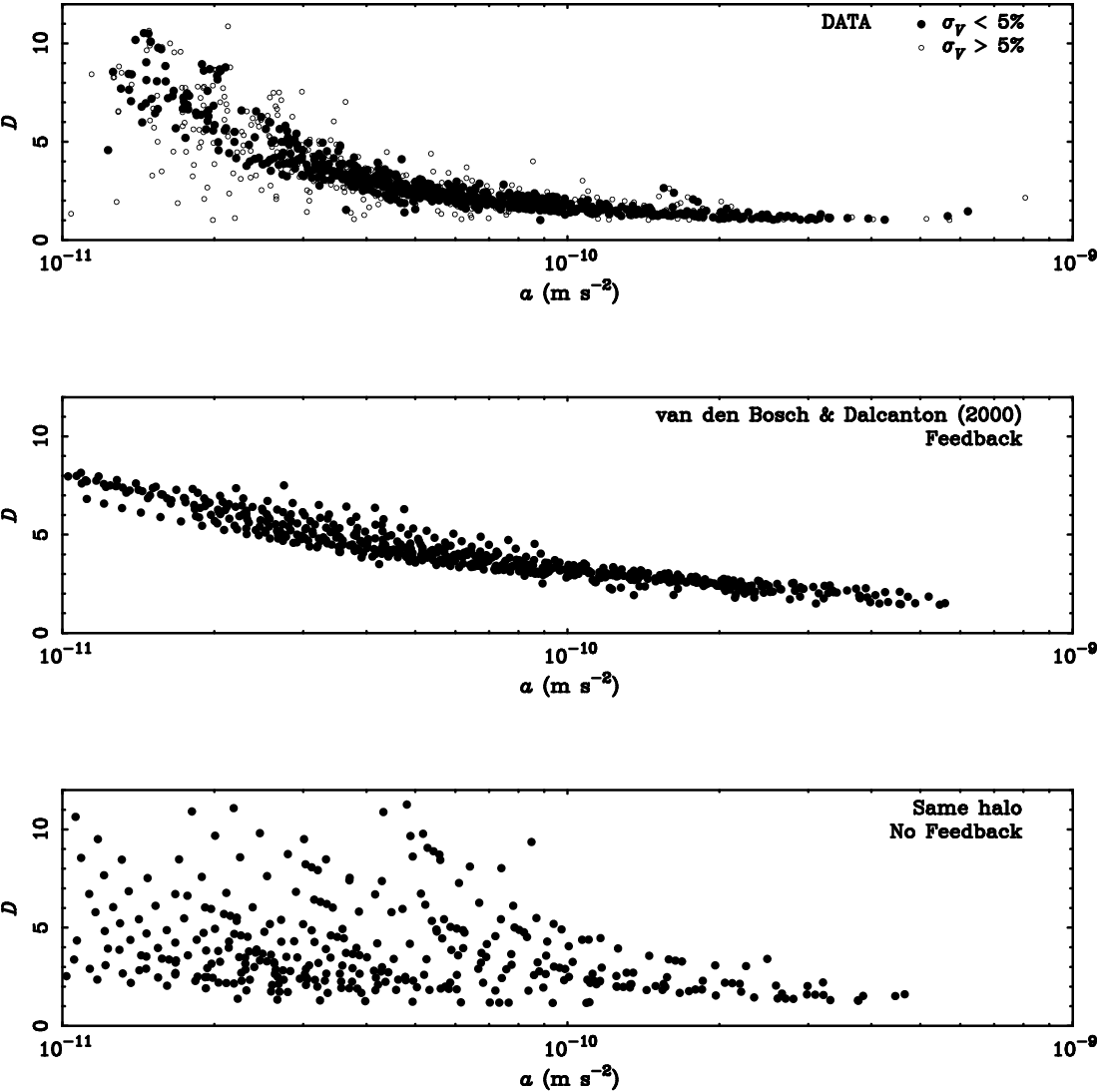}
      \caption{ The mass discrepancy ${\cal D} = V^2/\Vb^2$ correlates with
      acceleration $a = V^2/R$.  \textbf{Top panel:}  Data:  points with formal
      errors in velocity $< 5\%$ are shown as solid points, those with greater
      uncertainty as open points.
      The scatter is minimized (by construction) for MOND mass-to-light ratios,
      but the correlation persists for any plausible mass-to-light ratio 
      (McGaugh 2004).  \textbf{Middle panel:} 
      The feedback model of van den Bosch \& Dalcanton (2000), which 
      is to date the only attempt to quantitatively explain this aspect of the data
      with dark matter.  \textbf{Bottom panel:}
      The `same halo' model of McGaugh \& de Blok (1998a) illustrating
      the natural expectation for exponential disks residing in NFW halos sans
      feedback. }
       \label{MDdatanfw}
   \end{figure}

This must mean something.  The question, of course, is what.
There is a widespread myth that MOND is somehow designed to fit rotation
curves, and is guaranteed to do so.  This is demonstrably false (de Blok \&
McGaugh 1998).  Quite the contrary, MOND fits have only one fit parameter,
\ML, and are considerably better constrained than fits with dark halos.
This is a non-trivial fact: if you write down the wrong force law, it fails
(and fails badly) quite quickly.  Many other modifications of gravity have
been attempted, and suffered just such a fate.  That MOND works as well as
it does suggests that it is the correct \textit{effective}
force law, at least in spiral galaxies\footnote{There is not space to review
here the other successes of MOND, which are surprisingly numerous.  
That is not to say it is without problems.  The most serious at present,
to my mind, is the residual mass discrepancy in rich clusters of galaxies
(Aguirre, Schaye, \& Quataert 2001; Sanders 2003).  While it is tempting
to use this to dismiss the whole idea, on balance it is not obvious that
MOND is doing worse than CDM.}.  

We can rephrase MOND as a purely empirical relation (McGaugh 2005).
The amplitude of the mass discrepancy is correlated with acceleration
(Fig.~\ref{MDdatanfw}).  This mass discrepancy---acceleration relation
is present in the data, whether we call it MOND or not.  As such, it is an
important empirical fact which we need to understand.

I have devoted a good deal of time over the past decade trying to
understand this relation in conventional terms.  I have not succeeded.
There are a few claims by others to have done so, the most serious being
that of van den Bosch \& Dalcanton (2000).  Their model is plotted together
with the data in Fig.~\ref{MDdatanfw}, as is a model I would consider
natural\footnote{Van den Bosch \& Dalcanton (2000) describe
their model as natural.  I do not think this word means what they think
it means.} for CDM.  Many models are possible with dark matter,
which provides no unique null hypothesis:  there is a wide variety of
things rotation curves might reasonably be expected to do.
In MOND, there is precisely one thing rotation curves can do,
and that is what they do.
Philosophically, it is hard to imagine a stranger situation:
\textbf{why should dark matter look like MOND?}

\end{document}